\documentclass[journal]{IEEEtran}

\usepackage[T1]{fontenc}
\usepackage{amsmath,amssymb,amsfonts}
\usepackage{bm}
\usepackage{newtxtext,newtxmath}
\usepackage{graphicx}
\usepackage{booktabs}
\usepackage{algorithm}
\usepackage{algorithmic}
\usepackage{cite}
\usepackage{url}
\usepackage{color}
\usepackage{placeins}
\usepackage{float}
\usepackage{stmaryrd}
\usepackage{tabularx}
\usepackage{array}

\graphicspath{{figures/}}
\title{Online Wideband MIMO Channel Reconstruction from Periodically Swept RBs via Incremental CP Updates}

\author{Boxin Huang, Libin Zheng, Minru Bai, and Yuhao Jiang
\thanks{All authors are with Hunan University. This work was supported by the Sino-Russian Mathematics Challenge Fund (No.~2024SRMC07).}
}

\begin{document}

\maketitle

\begin{abstract}
Periodic resource-block (RB) scanning leaves most of the current wideband channel unobserved and mixes measurements of different ages. We develop an online canonical polyadic tracker with proximal block updates (CP-PBCD) that reconstructs the full channel after each narrow-RB acquisition. Age-weighted finite histories, frequency and temporal regularization, and bounded component management maintain an adaptive low-rank representation. Two warm-started conjugate-gradient iterations and parallel shifted frequency solves provide a fixed per-frame update budget. Training-user spatial projection strengthens noise suppression at low pilot SNR. Experiments cover ten test users, three speeds, four SNRs, and complete 1000-frame trajectories with 1-ms RB acquisition. Direct first-RB CP-PBCD achieves lower mean normalized mean squared error (NMSE) than all six baselines in all twelve conditions. At 3.6 km/h and 20 dB, it improves on periodic physical refitting by 3.91 dB and runs 35.1 times faster, averaging 0.717 ms per online update. Optional dispersed startup pilots improve early acquisition, while a controlled initialization study yields a last-100-frame NMSE range of 0.427 dB across pilot budgets. These results demonstrate accurate current-channel reconstruction through incremental updates with submillisecond average computation.
\end{abstract}


\section{Introduction}

\IEEEPARstart{A}{ccurate} channel state information (CSI) supports precoding and resource allocation in wideband massive multiple-input multiple-output (MIMO) systems \cite{marzetta2010noncooperative,larsson2014massive,lu2014overview}. We consider a pilot acquisition protocol in which the receiver observes one resource block (RB) per frame and periodically sweeps the band. This reduces the number of frequency-domain observations per frame but leaves most of the current channel unobserved. In our experiments, one of 48 RBs is observed every 1 ms, giving a 48-ms sweep period. After each observation, the receiver must reconstruct the current full-band channel using only the pilots acquired so far.

Periodic scanning couples incomplete observations with channel evolution: RBs stored during one sweep span 47 ms, but the required output represents the current instant. A receiver must therefore reuse historical frequency information while adapting to new measurements. Under a 1-ms acquisition interval, computation is also part of the reconstruction problem: an expensive update can delay delivery of the CSI needed for current transmission. We address this combination of narrow-band acquisition, unequal observation ages, and low-cost sequential reconstruction.

Tensor representations preserve the spatial, frequency, and temporal organization of CSI \cite{Kolda2009,Sidiropoulos2017,Cichocki2015}. Canonical polyadic (CP) models have been applied to mmWave MIMO-OFDM and dual-polarized channel estimation \cite{zhou2018low,liu2012tensor}, as well as intelligent reflecting surface-assisted estimation \cite{CE1}. Low-rank completion includes convex, factorized, and manifold-based approaches \cite{Gandy2011,Liu2013,kressner2014low,xu2013block}. Recovery analyses address random observations and incoherence \cite{yuan2017incoherent}, while structured sampling has also received direct attention. Ashraphijuo and Wang characterize deterministic conditions for finite and unique low-CP-rank completability \cite{ashraphijuo2017fundamental}, and Kanatsoulis et al. establish reconstruction from regular sub-Nyquist tensor samples \cite{kanatsoulis2020regular}. These studies characterize reconstruction from a sampling pattern. Periodically swept channel tracking adds a time-dependent output requirement: estimating current full-band CSI after each asynchronous RB acquisition.

Wan and Liu study hopping-pilot extrapolation through multi-band R-TST-MUSIC initialization followed by EM-based sparse Bayesian tracking of channel and synchronization parameters \cite{wan2024twostage}. Their two/four-BWP setting offers wider per-observation coverage than a single RB among 48. DDA-Net jointly reconstructs ten snapshots through Doppler--delay--angle unfolding, using 40-ms spacing and one of 17 frequency blocks \cite{ma2026dda}. Our setting requires a current full-band estimate after every 1-ms narrow-RB observation. We pursue this objective with a persistent CP representation and bounded incremental updates, and evaluate physical-model adaptations under the same acquisition stream.

Smoothness regularization and online tensor adaptation provide a basis for this task. Yokota et al. introduce total-variation and quadratic-variation penalties for smooth PARAFAC completion \cite{yokota2016smooth}. Adaptive PARAFAC updates decompositions as slices arrive \cite{nion2009adaptive}, streaming imputation handles incomplete tensors \cite{mardani2015subspace}, and OLSTEC applies recursive least squares to online CP tracking with incomplete observations \cite{kasai2016onlineICASSP}. Our formulation is tailored to the combination of periodic block visits, unequal observation ages, and a full-band output after every observed RB. It uses separate finite histories for the factor updates and age-aware weighting to control the contribution of stored observations. This organization complements recursive approaches such as OLSTEC, which accumulate exponentially weighted statistics.

The proposed online CP-PBCD tracker uses nonzero factor initialization, age-weighted histories, and bounded active-component management, followed by sequential temporal, spatial, and frequency proximal updates. The temporal step couples factor rows within a recent window, the spatial step combines recent observed slices, and the frequency step uses past visits to the current RB. Local difference penalties regularize the temporal window and observed frequency block. Their coupled normal equations have a block-tridiagonal structure. The GPU implementation updates the temporal window with two warm-started preconditioned conjugate-gradient (PCG) iterations and exploits a shared frequency Gram matrix for parallel shifted solves. The updated shared factors combine with stored frequency rows to reconstruct current full-band CSI after each observed RB. The experiments use one update sweep per frame.

Startup coverage determines how much frequency information is available before the first scan is complete. We compare first-RB initialization with a statistical initializer that adds six dispersed tones at the first frame. Small nonzero complementary components keep weak directions learnable, allowing subsequent observations to refine the representation. A controlled pilot-budget study separates improvements in early acquisition from the accuracy attained during later tracking.

The contributions are summarized as follows:
\begin{enumerate}
\item We formulate an online CP tracker for periodically swept RB observations. Separate finite histories and age-aware weights support sequential temporal, spatial, and frequency updates, followed by full-band reconstruction after each acquisition.
\item We derive coupled temporal-window and current-frequency-block subproblems and exploit their structure in a GPU implementation. Two warm-started PCG iterations, shared Gram computations, and parallel shifted solves reduce update cost. The implementation attains submillisecond average online updates on the reference workload.
\item We examine startup initialization under a fixed first RB with different numbers of additional dispersed pilots. Nonzero component completion removes a zero-state limitation, and improved startup coverage mainly benefits early acquisition; late tracking errors are substantially closer across the tested initial estimates.
\item We evaluate direct first-RB and six-extra-pilot versions over ten users, three speeds, and four SNRs, and achieve lower full-trajectory mean NMSE than six baselines in all twelve conditions with first-RB CP-PBCD. A two-by-two ablation identifies the contribution of temporal regularization.
\end{enumerate}

\section{Online Channel Estimation Problem Formulation}

We represent the channel as a tensor and fit a low-rank CP model to the observed RBs. The online updates use separate observation histories, with local difference penalties on the frequency and temporal factors.

\subsection{Channel Tensor and Block Sampling Model}

Consider a wideband OFDM-based massive MIMO system with frequency-domain channel response $\mathbf h_{f,t}$ at subcarrier $f$ and frame $t$. Stacking the spatial entries (transmit/receive antenna pairs or beams), subcarriers, and frames gives a third-order complex channel tensor:

\begin{equation}
\mathcal{H} \in \mathbb{C}^{N_T \times N_F \times N_t},
\quad
\mathcal{H}_{:,f,t} = \mathbf{h}_{f,t},
\end{equation}

where $N_T$ is the spatial dimension (e.g., the number of antenna pairs or beams), $N_F$ is the number of subcarriers, and $N_t$ is the number of frames.

Each observed RB contains $s$ consecutive subcarriers. At frame $t$, the index set $\mathcal I_t$ selects the current block, and periodic shifts visit the entire band.

Let $\mathbf{S}_t$ denote the observed slice of $\mathcal{H}$ at time $t$. The observation model is given by

\begin{equation}
\mathbf{S}_t =
\mathcal{P}_{\mathcal{I}_t}
\big(
\mathcal{H}_{:,:,t}
\big)
+
\mathbf{E}_t
\in \mathbb{C}^{N_T \times s},
\label{eq:observation_model}
\end{equation}

where $\mathcal{P}_{\mathcal{I}_t}$ is a frequency selection operator retaining columns indexed by $\mathcal{I}_t$, and $\mathbf{E}_t$ represents additive white Gaussian noise.

\begin{figure}[t]
    \centering
    \includegraphics[width=0.86\linewidth]{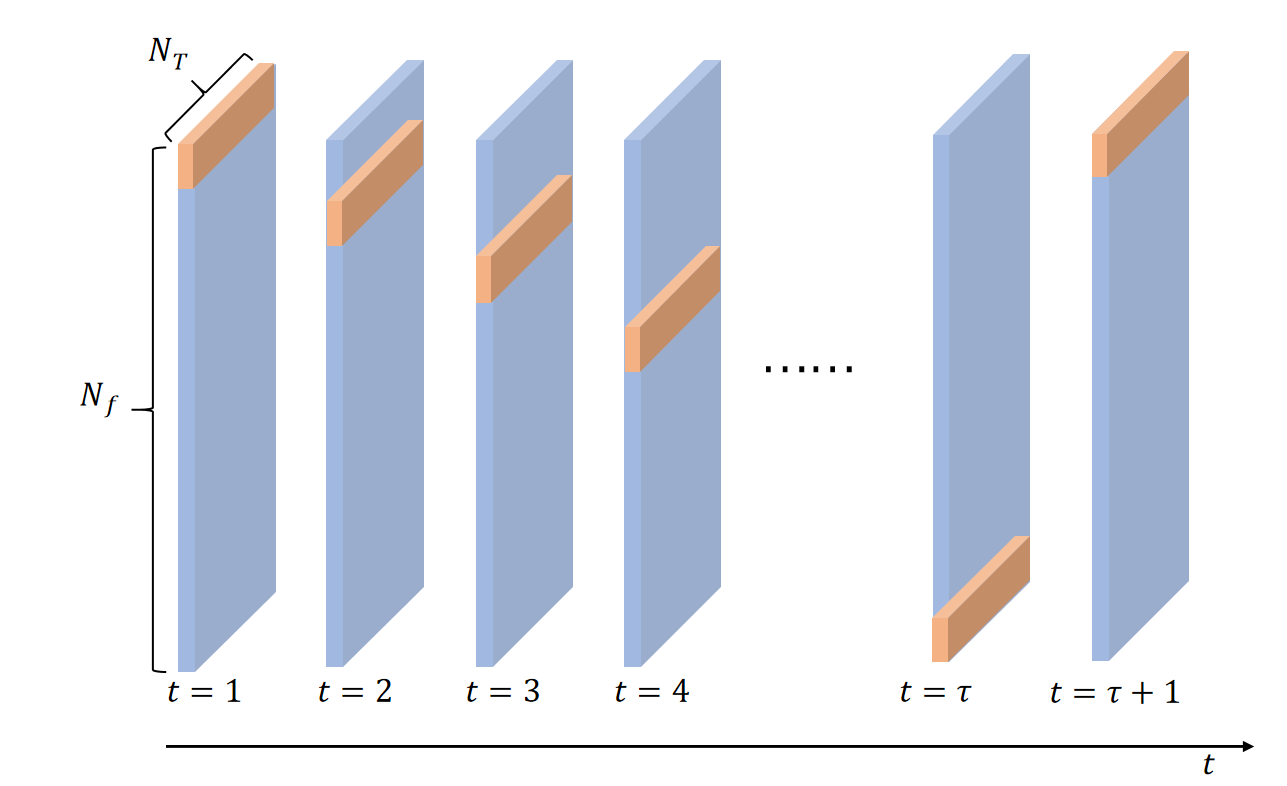}
    \caption{Illustration of the adopted RB-inspired periodically swept block-sampling pattern. Contiguous RB observations shift regularly across time frames, preserving local frequency correlation while covering different frequency regions over time.}
    \label{fig:block_sample}
\end{figure}

The sampling pattern selects contiguous frequency blocks and shifts them regularly across successive frames, as shown in Fig.~\ref{fig:block_sample}. It preserves local frequency correlation, unlike independent random-entry sampling.

At startup, $\Omega_0=\mathcal I_0\cup\mathcal E_0$ adds dispersed tones to the first RB. Spatial and temporal histories use all $s+|\mathcal E_0|$ measured columns, with zero weights on batch padding. The frequency history retains the complete first RB. Subsequent observations follow the ordinary scan.

\subsection{Low-Rank Channel Modeling via CP Decomposition}

The spatial, frequency, and temporal sparsity of wideband MIMO channels motivates a low-rank tensor approximation.

We model $\mathcal{H}$ using a rank-$R$ CP (Canonical Polyadic) decomposition \cite{Hitchcock1927,Carroll1970,Harshman1970,sidiropoulos2000parallel}:

\begin{equation}
\mathcal{H}
\approx
\llbracket
\mathbf{U}_T,
\mathbf{U}_F,
\mathbf{U}_t
\rrbracket
=
\sum_{r=1}^{R}
\mathbf{u}_{T,r}
\circ
\mathbf{u}_{F,r}
\circ
\mathbf{u}_{t,r},
\label{eq:cp_model}
\end{equation}

where $\circ$ denotes the vector outer product. The factor matrices are defined as

\begin{align}
\mathbf{U}_T &= [\mathbf{u}_{T,1},\dots,\mathbf{u}_{T,R}]
\in \mathbb{C}^{N_T \times R}, \\
\mathbf{U}_F &= [\mathbf{u}_{F,1},\dots,\mathbf{u}_{F,R}]
\in \mathbb{C}^{N_F \times R}, \\
\mathbf{U}_t &= [\mathbf{u}_{t,1},\dots,\mathbf{u}_{t,R}]
\in \mathbb{C}^{N_t \times R}.
\end{align}

The observed slice follows directly from the CP model:
\begin{equation}
\mathbf S_t^\star=\mathbf U_T\operatorname{diag}(\mathbf U_t[t,:])\mathbf U_F[\mathcal I_t,:]^T.
\label{eq:slice_equivalence}
\end{equation}
The measured slice adds noise and the sampled CP approximation residual.

\subsection{Physical Motivation and Denoising Role of Smoothness Regularization}

Limited delay and Doppler spreads create local frequency and temporal correlation. First-order penalties on neighboring factor rows exploit this continuity and suppress rapid noise-induced fluctuations. We apply them within the current RB and temporal history, using fixed factor conventions and explicit regularization weights.

\subsection{Block-Local Regularized Online Estimation}

The estimator maintains three observation histories. Let $\mathcal A_t$ contain the most recent $L_A$ observed frames, $\mathcal C_t$ the most recent $L_C$ observed frames, and $\mathcal V_t$ the most recent $K_F$ visits to the current frequency block $\mathcal I_t$. At startup, each history contains only the observations available so far. The main experiments use $L_A=48$, $L_C=24$, and $K_F=3$, with one observed RB ($s=12$ subcarriers) per frame. Visits in $\mathcal V_t$ are generally nonconsecutive. For the 48-RB sweep, successive visits are 48 frames apart.

Frequency regularization couples adjacent rows within the observed block, and temporal regularization couples consecutive rows within $\mathcal C_t$. Both penalties stop at the boundaries of their updated blocks. Unobserved frequency rows retain their stored values; full-band reconstruction combines them with the updated shared spatial factor and current temporal row.

The three proximal least-squares subproblems below define CP-PBCD. The smoothness weights $\tau_F,\tau_t\geq0$ may be set to zero in ablations. Positive proximal weights $\gamma_T,\gamma_F,\gamma_t>0$ keep each local linear system nonsingular, with proximal centers taken from the factor values immediately before each update.

\section{Solution Algorithm: Sequential Proximal Block Updates}

At every observed frame, the update order is temporal, spatial, and then frequency. A superscript $-$ denotes the value immediately before a subproblem, and $+$ its solution; other factors in that subproblem take their latest available values. All $R$ components of each updated row are coupled in the update. The equations below define the regularized subproblems; the main GPU implementation uses the fixed-budget solver specified below. The main experiments perform exactly one sweep per frame, without a convergence-based stopping test. At $t>0$, the new temporal row is copied from the preceding row once before the first temporal update; the first frame uses the chosen initializer.

\begin{figure}[!htb]
\centering
\includegraphics[width=\linewidth]{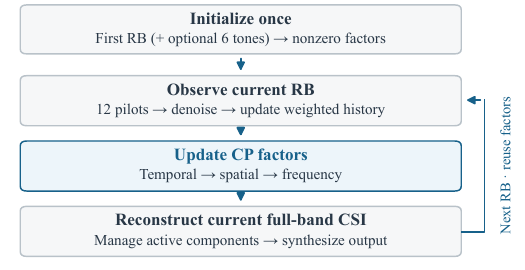}
\caption{Online reconstruction workflow. Initialization runs once; subsequent RB acquisitions reuse the CP factors and update the current full-band channel.}
\label{fig:overall_pipeline}
\end{figure}

Figure~\ref{fig:overall_pipeline} summarizes the startup and recurring per-frame operations.

\subsection{Temporal-Factor Update}

Write $\mathcal C_t=(c_1,\ldots,c_m)$ in chronological order and $\mathcal I_{c_i}=(f_{i,1},\ldots,f_{i,s})$. Define $\mathbf x_i=\mathbf U_t[c_i,:]^T\in\mathbb C^R$ and $\mathbf y_i=\operatorname{vec}(\mathbf S_{c_i}^T)\in\mathbb C^{N_Ts}$, where $\operatorname{vec}$ stacks columns. The design matrix $\mathbf D_i\in\mathbb C^{N_Ts\times R}$ has entries
\begin{equation}
\mathbf D_i[(n-1)s+j,r]
=\mathbf U_T[n,r]\mathbf U_F[f_{i,j},r].
\label{eq:temporal_design}
\end{equation}
This ordering matches the row-wise flattening of the observed spatial-by-frequency slice. The temporal step solves
\begin{equation}
\begin{aligned}
\min_{\mathbf x_1,\ldots,\mathbf x_m}\;&
\frac12\sum_{i=1}^m\|\mathbf y_i-\mathbf D_i\mathbf x_i\|_2^2\\
&+\frac{\tau_t}{2}\sum_{i=1}^{m-1}\|\mathbf x_{i+1}-\mathbf x_i\|_2^2\\
&+\frac{\gamma_t}{2}\sum_{i=1}^m\|\mathbf x_i-\mathbf x_i^-\|_2^2.
\end{aligned}
\label{eq:Ut_subproblem}
\end{equation}
The solution replaces only the rows indexed by $\mathcal C_t$. The penalty stops at the window boundary, with no term linking $\mathbf x_1$ to an earlier stored row. Copying the preceding row to initialize a new row sets its proximal center but adds no boundary term to \eqref{eq:Ut_subproblem}.

\subsection{Spatial-Factor Update}

Using the newly updated temporal factor, define, for $\ell\in\mathcal A_t$,
\[
\mathbf Q_\ell=\operatorname{diag}(\mathbf U_t[\ell,:])
\mathbf U_F[\mathcal I_\ell,:]^T\in\mathbb C^{R\times s}.
\]
The spatial step is
\begin{equation}
\min_{\mathbf U_T}\;
\frac12\sum_{\ell\in\mathcal A_t}\|\mathbf S_\ell-\mathbf U_T\mathbf Q_\ell\|_F^2
+\frac{\gamma_T}{2}\|\mathbf U_T-\mathbf U_T^-\|_F^2,
\label{eq:UT_subproblem}
\end{equation}
with solution
\begin{equation}
\mathbf U_T^+=
\left(\sum_{\ell\in\mathcal A_t}\mathbf S_\ell\mathbf Q_\ell^H+\gamma_T\mathbf U_T^-\right)
\left(\sum_{\ell\in\mathcal A_t}\mathbf Q_\ell\mathbf Q_\ell^H+\gamma_T\mathbf I_R\right)^{-1}.
\label{eq:UT_update}
\end{equation}
The implementation evaluates the equivalent transposed linear solve. Temporal rows in $\mathcal A_t$ but outside $\mathcal C_t$ retain their stored values.

\subsection{Current-Block Frequency-Factor Update}

Let $\mathcal I_t=(f_1,\ldots,f_s)$ and $\mathcal V_t=(v_1,\ldots,v_k)$. For each $f_j$, define
\begin{equation}
\begin{aligned}
\mathbf A_j&=\begin{bmatrix}
\mathbf U_T\operatorname{diag}(\mathbf U_t[v_1,:])\\
\vdots\\
\mathbf U_T\operatorname{diag}(\mathbf U_t[v_k,:])
\end{bmatrix},&
\mathbf z_j&=\begin{bmatrix}\mathbf S_{v_1}[:,j]\\\vdots\\\mathbf S_{v_k}[:,j]\end{bmatrix},
\end{aligned}
\label{eq:frequency_design}
\end{equation}
where $\mathbf A_j\in\mathbb C^{kN_T\times R}$. All visits refer to the same frequency block. These design matrices are identical across $j$, so the GPU implementation forms their common Gram matrix once. With $\mathbf b_j=\mathbf U_F[f_j,:]^T$, the step solves
\begin{equation}
\begin{aligned}
\min_{\mathbf b_1,\ldots,\mathbf b_s}\;&
\frac12\sum_{j=1}^s\|\mathbf z_j-\mathbf A_j\mathbf b_j\|_2^2\\
&+\frac{\tau_F}{2}\sum_{j=1}^{s-1}\|\mathbf b_{j+1}-\mathbf b_j\|_2^2\\
&+\frac{\gamma_F}{2}\sum_{j=1}^s\|\mathbf b_j-\mathbf b_j^-\|_2^2.
\end{aligned}
\label{eq:UF_subproblem}
\end{equation}
The solution replaces only $\mathbf U_F[\mathcal I_t,:]$ and contains $sR$ unknowns spanning all components of the observed block. No penalty links $f_1$ or $f_s$ to subcarriers outside $\mathcal I_t$; the remaining frequency rows are stored unchanged.

\subsection{Joint Block-Tridiagonal Solver}

Both \eqref{eq:Ut_subproblem} and \eqref{eq:UF_subproblem} have the generic form
\begin{equation}
\begin{aligned}
\min_{\{\mathbf x_i\}_{i=1}^q}\;&\frac12\sum_{i=1}^q\|\mathbf y_i-\mathbf B_i\mathbf x_i\|_2^2
+\frac{\gamma}{2}\sum_{i=1}^q\|\mathbf x_i-\mathbf x_i^-\|_2^2\\
&+\frac{\tau}{2}\sum_{i=1}^{q-1}\|\mathbf x_{i+1}-\mathbf x_i\|_2^2.
\end{aligned}
\label{eq:block_template}
\end{equation}
Let $d_i$ be the number of neighbors of row $i$ within this chain: $d_1=d_q=1$ and $d_i=2$ for interior rows when $q>1$; $d_1=0$ when $q=1$. Set
\begin{equation}
\begin{aligned}
\mathbf G_i&=\mathbf B_i^H\mathbf B_i+(\gamma+\tau d_i)\mathbf I_R,\\
\mathbf r_i&=\mathbf B_i^H\mathbf y_i+\gamma\mathbf x_i^-.
\end{aligned}
\label{eq:block_diagonal_rhs}
\end{equation}
The normal equations are block tridiagonal:
\begin{equation}
\mathbf G_i\mathbf x_i
-\tau\!\!\sum_{j\in\{i-1,i+1\}\cap\{1,\ldots,q\}}\!\!\mathbf x_j
=\mathbf r_i,\quad i=1,\ldots,q.
\label{eq:block_normal_equations}
\end{equation}
Each diagonal block is $R\times R$, generally dense because the CP components are coupled through $\mathbf B_i^H\mathbf B_i$. Each off-diagonal block is $-\tau\mathbf I_R$. The complete system is Hermitian positive definite for $\gamma>0$. 

The block Thomas elimination can be written as
\begin{equation}
\begin{aligned}
\mathbf P_1&=\mathbf G_1,&\mathbf h_1&=\mathbf r_1,\\
\mathbf P_i&=\mathbf G_i-\tau^2\mathbf P_{i-1}^{-1},&
\mathbf h_i&=\mathbf r_i+\tau\mathbf P_{i-1}^{-1}\mathbf h_{i-1},
\end{aligned}
\label{eq:block_forward}
\end{equation}
for $i=2,\ldots,q$, followed by
\begin{equation}
\begin{aligned}
\mathbf x_q&=\mathbf P_q^{-1}\mathbf h_q,\\
\mathbf x_i&=\mathbf P_i^{-1}(\mathbf h_i+\tau\mathbf x_{i+1}),
\quad i=q-1,\ldots,1.
\end{aligned}
\label{eq:block_backward}
\end{equation}
The block elimination above defines an exact reference solver for the regularized subproblems. The temporal step uses $(q,\tau,\gamma)=(m,\tau_t,\gamma_t)$, and the frequency step uses $(s,\tau_F,\gamma_F)$. The main experiments use the following GPU realization of these same systems.

\subsection{GPU Implementation with a Fixed Update Budget}
Let $\mathbf G_i=\mathbf B_i^H\mathbf B_i$ include the age weight of row $i$, and let $\mathbf L_q=\mathbf Q_q\boldsymbol\Lambda_q\mathbf Q_q^T$ be the path-graph Laplacian for the first-difference penalty. The temporal normal matrix is
\[
\mathbf K=\operatorname{blkdiag}(\mathbf G_1,\ldots,\mathbf G_q)
+\gamma\mathbf I_{qR}+\tau\mathbf L_q\otimes\mathbf I_R.
\]
We use the positive-definite preconditioner
\[
\mathbf M=\mathbf I_q\otimes(\overline{\mathbf D}+\gamma\mathbf I_R)
+\tau\mathbf L_q\otimes\mathbf I_R,
\quad
\overline{\mathbf D}=q^{-1}\sum_i\operatorname{Diag}(\operatorname{diag}\mathbf G_i).
\]
Applying $\mathbf M^{-1}$ requires transforms by $\mathbf Q_q$ and elementwise division by $\overline D_{rr}+\gamma+\tau\Lambda_{\ell\ell}$. Starting from the stored temporal rows, the implementation performs exactly two PCG iterations. This gives a bounded solve budget for each arriving frame.

For the frequency update, all rows share $\mathbf G$. Transforming along the $s$ frequency rows diagonalizes their difference penalty and yields $s$ independent systems with matrices $\mathbf G+(\gamma_F+\tau_F\Lambda_{\ell\ell})\mathbf I_R$. These dense systems are solved in parallel and transformed back. The spatial step uses a dense multiple-right-hand-side solve. Batched Gram operations reuse the CP product structure, and CUDA Graph replay reduces launch overhead. Frequency smoothness, temporal smoothness, age weighting, and active-component checks remain enabled in the main experiments.

\begin{algorithm}[t]
\caption{Implemented online CP-PBCD update}
\label{alg:CP_PBCD}
\begin{algorithmic}[1]
\REQUIRE Observed slices $\mathbf S_t$, blocks $\mathcal I_t$, nonzero initialized factors, history limits $L_A,L_C,K_F$, weights, fixed sweep count $I$
\STATE For statistical startup, preload frame 0 and output its initialized estimate as in Section~\ref{sec:initialization}.
\FOR{each observed frame $t$, starting at 0 for direct first-RB initialization or 1 for statistical startup}
\STATE Append observations to spatial and temporal histories; retain at most $L_A$ and $L_C$ frames.
\STATE Append $t$ to the visit history of $\mathcal I_t$; retain at most $K_F$ visits for that block.
\IF{$t>0$}
\STATE Set $\mathbf U_t[t,:]\gets\mathbf U_t[t-1,:]$ once.
\ENDIF
\STATE Compute the normalized age weights for each retained history.
\FOR{$i=1,\ldots,I$}
\STATE Update $\mathbf U_t[\mathcal C_t,:]$ for \eqref{eq:Ut_subproblem} with two warm-started PCG iterations.
\STATE Update $\mathbf U_T$ by \eqref{eq:UT_update}, using $\mathcal A_t$.
\STATE Update $\mathbf U_F[\mathcal I_t,:]$ by the parallel shifted solves for \eqref{eq:UF_subproblem}.
\ENDFOR
\STATE Perform the scheduled active-component check and update the factors.
\STATE Output $\widehat{\mathbf H}_t=\mathbf U_T\operatorname{diag}(\mathbf U_t[t,:])\mathbf U_F^T$.
\ENDFOR
\end{algorithmic}
\end{algorithm}

\subsection{Age-Aware History Weighting}
Older observations can become less representative of the current factors as the channel evolves. The main implementation therefore assigns exponentially decreasing weights within each retained history. For $\mathcal H\in\{\mathcal A_t,\mathcal C_t,\mathcal V_t\}$, the weight of entry $k$ at frame $t$ is
\begin{equation}
w_{t,k}^{(h)}=\frac{2^{-(t-k)/h}}{|\mathcal H|^{-1}\sum_{j\in\mathcal H}2^{-(t-j)/h}},
\label{eq:age_weights}
\end{equation}
with half-lives $h=48,24,96$ frames for spatial, temporal, and frequency histories, respectively. Ages use elapsed frame indices, including the gaps between frequency-block visits. The normalization gives $|\mathcal H|^{-1}\sum_{k\in\mathcal H}w_{t,k}^{(h)}=1$: it redistributes the data weights toward recent observations without reducing their mean scale relative to the unchanged penalties.

To express how these weights enter a block update, let $\boldsymbol x$ collect its unknown factor entries, and let $\boldsymbol y_k$ and $\mathbf D_k$ denote the observation vector and design matrix for history entry $k$, with the other factors fixed. For a joint temporal update, $\mathbf D_k$ includes the selection of the corresponding temporal row. The weighted quadratic subproblem is
\begin{equation}
\min_{\boldsymbol x}\;
\frac12\sum_{k\in\mathcal H}w_{t,k}^{(h)}
\|\boldsymbol y_k-\mathbf D_k\boldsymbol x\|_2^2
+\mathcal R_t(\boldsymbol x),
\label{eq:age_subproblem}
\end{equation}
where $\mathcal R_t$ contains the original proximal and difference penalties. Writing their contribution as $\mathcal R_t(\boldsymbol x)=\frac12\boldsymbol x^H\mathbf Q_t\boldsymbol x-\operatorname{Re}(\boldsymbol b_t^H\boldsymbol x)+\mathrm{const}$ yields
\begin{equation}
\begin{split}
\left(\sum_{k\in\mathcal H}w_{t,k}^{(h)}\mathbf D_k^H\mathbf D_k
+\mathbf Q_t\right)\boldsymbol x
\\
=\sum_{k\in\mathcal H}w_{t,k}^{(h)}\mathbf D_k^H\boldsymbol y_k
+\boldsymbol b_t.
\end{split}
\label{eq:age_normal}
\end{equation}
Thus, each data Gram matrix and right-hand-side contribution is multiplied by the same age weight, while $\mathbf Q_t$ and $\boldsymbol b_t$ are unchanged. Equivalently, $\mathbf D_k$ and $\boldsymbol y_k$ are both scaled by $\sqrt{w_{t,k}^{(h)}}$. The spatial and frequency steps solve their resulting systems, whereas the temporal step applies the fixed two-PCG-iteration budget described above. Age weighting changes the relative influence of retained observations; active-component selection is handled separately below.

\subsection{Bounded Active-Component Management}
The online implementation adapts the number of active CP components through bounded deactivation and restoration of existing factor columns. Let $\mathcal J_t$ denote the active column indices and $r_t$ their number:
\begin{equation}
\mathcal J_t\subseteq\{1,\ldots,R_{\max}\},\qquad
r_t=|\mathcal J_t|,\qquad R_{\min}\le r_t\le R_{\max},
\label{eq:active_set}
\end{equation}
where $R_{\min}=12$, $R_{\max}=40$, and all 40 columns are initially active. 

Checks occur at $t=95,143,191,\ldots$, after the factor updates. Let $\mathcal W_t=\{t-47,\ldots,t\}$, and stack the noisy observed blocks $\mathbf S_k$, $k\in\mathcal W_t$, into $\boldsymbol y_t^{\mathrm{obs}}$. Denote the current spatial and frequency columns by $\boldsymbol a_r=\mathbf U_T[:,r]$ and $\boldsymbol b_r=\mathbf U_F[:,r]$, and the stored temporal coefficient by $c_{k,r}=\mathbf U_t[k,r]$. The contribution of component $r$ on this window is
\begin{equation}
\begin{split}
\boldsymbol u_{r,t}
&=\operatorname{col}_{k\in\mathcal W_t}
\operatorname{vec}\!\left(c_{k,r}\boldsymbol a_r
\boldsymbol b_r[\mathcal I_k]^T\right),\\
\boldsymbol e_t
&=\boldsymbol y_t^{\mathrm{obs}}-
\sum_{r\in\mathcal J_t}\boldsymbol u_{r,t}.
\end{split}
\label{eq:component_residual}
\end{equation}
Here $\operatorname{col}$ stacks vectors in time order. Define $E_t=\max\{\|\boldsymbol y_t^{\mathrm{obs}}\|_2^2,\epsilon\}$ with the implementation safeguard $\epsilon=10^{-30}$. Component selection uses these unweighted observed-block energies, independently of the age weights in the factor updates.

\emph{Deactivation:} The relative contribution energy and the relative increase in squared error upon removal are
\begin{equation}
\begin{split}
q_{r,t}&=\frac{\|\boldsymbol u_{r,t}\|_2^2}{E_t},\\
d_{r,t}&=\frac{\|\boldsymbol e_t+\boldsymbol u_{r,t}\|_2^2
-\|\boldsymbol e_t\|_2^2}{E_t}.
\end{split}
\label{eq:component_drop_scores}
\end{equation}
A component becomes a removal candidate only after two consecutive low-contribution checks. It is deactivated if
\begin{equation}
q_{r,t}<10^{-3},\qquad q_{r,t-48}<10^{-3},\qquad
d_{r,t}\le10^{-4}.
\label{eq:component_drop_rule}
\end{equation}
No removal occurs at the first check, since a previous low-contribution check is required. Candidates are examined in increasing contribution energy. After each accepted removal, the residual is updated as $\boldsymbol e_t\gets\boldsymbol e_t+\boldsymbol u_{r,t}$ before testing the next candidate. At most two components are removed per check, and $r_t$ cannot fall below $R_{\min}$. The current spatial and frequency columns are cached with the deactivation frame; their live columns and all stored temporal coefficients are then set to zero.

\emph{Restoration:} For an eligible cached pair $(\overline{\boldsymbol a}_r,\overline{\boldsymbol b}_r)$, define the direction
\[
\boldsymbol v_{r,t}=\operatorname{col}_{k\in\mathcal W_t}
\operatorname{vec}\!\left(\overline{\boldsymbol a}_r
\overline{\boldsymbol b}_r[\mathcal I_k]^T\right).
\]
Using the residual after any removals, the common-coefficient fit and its normalized gain score are
\begin{equation}
\begin{split}
\nu_{r,t}&=\max\{\|\boldsymbol v_{r,t}\|_2^2,\epsilon\},\\
\alpha_{r,t}&=\frac{\boldsymbol v_{r,t}^H\boldsymbol e_t}{\nu_{r,t}},
\qquad
g_{r,t}=\frac{|\boldsymbol v_{r,t}^H\boldsymbol e_t|^2}{\nu_{r,t}E_t}.
\end{split}
\label{eq:component_restore_score}
\end{equation}
When $\|\boldsymbol v_{r,t}\|_2^2\ge\epsilon$, $\alpha_{r,t}$ minimizes $\|\boldsymbol e_t-\alpha\boldsymbol v_{r,t}\|_2^2$, and $g_{r,t}$ is the resulting squared-residual reduction divided by $E_t$. A dormant component is eligible after a 96-frame cooldown and only if $g_{r,t}>10^{-3}$. The eligible component with the largest gain is restored, at most one per check. Its cached spatial and frequency columns are reinstated, and $c_{k,r}$ is set to $\alpha_{r,t}$ for $k\in\mathcal W_t$.

Observed-block residuals drive both decisions. Cached directions allow restoration, and the solver retains its allocated width of 40 for stable GPU execution.

\section{Computational Complexity}

We count the arithmetic operations in the GPU implementation, with $N_T$ spatial channels, $s$ observed subcarriers, allocated width $R$, and history limits $L_A,L_C,K_F$. The spatial step forms an $R\times R$ normal matrix and solves for $N_T$ right-hand sides:
\begin{equation}
C_A=O(L_AsN_TR+L_AsR^2+N_TR^2+R^3).
\label{eq:cut_complexity}
\end{equation}
The frequency step forms one shared Gram matrix and $s$ right-hand sides. Diagonalizing the fixed difference operator leaves $s$ parallel shifted systems:
\begin{equation}
C_F=O(K_FN_TR^2+K_FN_TsR+sR^3+s^2R).
\label{eq:cuf_complexity}
\end{equation}
For the temporal step, the CP design Gram is a Hadamard product of spatial and observed-frequency Grams. A PCG iteration costs $O(L_CR^2+L_C^2R)$, including the preconditioner transforms. Hence,
\begin{equation}
\begin{split}
C_C=O\big(&N_TR^2+L_CsR^2+L_CN_TsR\\
&+J(L_CR^2+L_C^2R)\big),\qquad J=2.
\end{split}
\label{eq:cutime_complexity}
\end{equation}
With $I=1$ update sweep and full-band synthesis, the per-frame cost is
\begin{equation}
C_{\mathrm{frame}}=O\!\left(I(C_A+C_F+C_C)+N_TN_FR\right).
\label{eq:cppbcd_complexity}
\end{equation}
The fixed Laplacian bases are prepared once. Active-component checks add periodic observed-history energy and residual computations; they are included in the runtime measurements below. Initialization, statistical-prior preparation, and GPU graph preparation are separate setup costs. The exact block-elimination reference costs $O(qR^3)$ after assembly, whereas the deployed temporal solver uses the fixed two-iteration budget. Full-band synthesis remains explicit in the cost above.

\section{Local Tracking Analysis}

We analyze how observation perturbations and reference mismatch propagate through the online updates on a fixed active set. Under the local contraction condition stated below, the analysis relates the accumulated perturbation to the long-run tracking error.

Let $\Theta_t$ collect the spatial and frequency factors together with the stored temporal rows needed by the three histories. Write the implemented frame update as $\Theta_t=\mathcal F_t(\Theta_{t-1})$. The map includes history advancement, copying the preceding temporal row, two temporal PCG iterations, and the fixed number of temporal--spatial--frequency sweeps. Histories are reindexed consistently when comparing states. Let $\mathcal F_t^0$ be the corresponding update using noiseless observations and let $\Theta_t^\star$ be a chosen bounded reference trajectory in a fixed CP scaling and permutation convention. 

Define the state error and reference defect by
\begin{equation}
\begin{aligned}
e_t&=\|\Theta_t-\Theta_t^\star\|,\\
b_t&=\|\mathcal F_t^0(\Theta_{t-1}^\star)-\Theta_t^\star\|.
\end{aligned}
\label{eq:reference_defect}
\end{equation}
The defect measures how closely the chosen channel-factor reference follows the actual noiseless algorithm. It can contain temporal drift, finite-rank mismatch, regularization bias, and the fixed-iteration solve approximation. 

\textbf{Assumption 1 (Bounded local evolution).}
The actual iterates, reference states, and intermediate noiseless trajectories remain in a common bounded neighborhood. In this neighborhood each $\mathcal F_t^0$ is Lipschitz with a uniform constant $M<\infty$.

\textbf{Assumption 2 (Bounded update perturbation).}
For states in this neighborhood, the observation-induced perturbation satisfies
\[
\|\mathcal F_t(\Theta)-\mathcal F_t^0(\Theta)\|\leq\eta_t,
\qquad \sup_t\mathbb E(\eta_t+b_t)\leq B<\infty.
\]
The update-level perturbation includes all overlapping histories.

\textbf{Assumption 3 (Contraction over a coverage period).}
Every RB is visited during a period of $W$ frames, and the noiseless composite update
\[
\mathcal G_t^0=\mathcal F_{t+W}^0\circ\cdots\circ\mathcal F_{t+1}^0
\]
is Lipschitz with a uniform constant $\gamma<1$ in the stated neighborhood. 

\textbf{Theorem 1 (Conditional local tracking bound).}
Under Assumptions 1--3 and finite initial error,
\[
\mathbb E e_{t+kW}\leq\gamma^k\mathbb E e_t+
\frac{C_W B(1-\gamma^k)}{1-\gamma},
\]
and consequently
\begin{equation}
\limsup_{t\to\infty}\mathbb E e_t
\leq \frac{C_W B}{1-\gamma},
\label{eq:tracking_error_bound}
\end{equation}
where $C_W$ depends on $W$ and the uniform one-frame Lipschitz bound.

\begin{IEEEproof}
Set $a_t=\eta_t+b_t$. Over $W$ steps, compare the actual trajectory with the noiseless trajectory starting from $\Theta_t$, and compare the reference trajectory with the noiseless trajectory starting from $\Theta_t^\star$. Repeated use of the one-frame Lipschitz bound controls the accumulated perturbations. Applying the composite contraction to the two noiseless trajectories gives
\begin{equation}
e_{t+W}\leq\gamma e_t+\sum_{j=1}^W M^{W-j}a_{t+j}.
\label{eq:block_recursion}
\end{equation}
Taking expectations and writing $C_W=\sum_{j=1}^W M^{W-j}$ yields
\[
\mathbb E e_{t+W}\leq\gamma\mathbb E e_t+C_W B.
\]
Iteration for each of the $W$ possible starting offsets proves \eqref{eq:tracking_error_bound}. Assumption 1 keeps the compared trajectories within the common neighborhood.
\end{IEEEproof}

\textbf{Corollary 1 (Current-frame reconstruction bound).}
Let $\mathcal R_t(\Theta)=\mathbf U_T\operatorname{diag}(\mathbf U_t[t,:])\mathbf U_F^T$ and let
\[
\varepsilon_t^\star=\|\mathcal R_t(\Theta_t^\star)-\mathbf H_t^\star\|_F
\]
be the mismatch of the chosen reference to the noiseless current channel. On the bounded factor set, the synthesis map is Lipschitz, so
\begin{equation}
\limsup_t\mathbb E\|\widehat{\mathbf H}_t-\mathbf H_t^\star\|_F
\leq\frac{C_{\mathrm{rec}}B}{1-\gamma}+\limsup_t\varepsilon_t^\star.
\label{eq:reconstruction_error_bound}
\end{equation}
\begin{IEEEproof}
Bounded factor norms make the polynomial CP synthesis map locally Lipschitz, and hence
\[
\|\widehat{\mathbf H}_t-\mathbf H_t^\star\|_F
\leq C_{\mathrm{CP}}e_t+\varepsilon_t^\star.
\]
Combining this inequality with the state bound proves Corollary~1.
\end{IEEEproof}

The factor $\gamma^k$ describes attenuation of initialization error across coverage periods, while $C_WB/(1-\gamma)$ describes accumulated observation and model perturbations. This distinction motivates the separate evaluation of startup quality and late tracking accuracy.

\section{Initialization and Startup Pilot Coverage}
\label{sec:initialization}
The first RB supplies 12 adjacent tones for a 576-subcarrier output. Both initializers maintain 40 learnable components; they differ in the frequency coverage available at startup.

\subsection{Direct First-RB Initialization}
Let $\mathbf Y_0\in\mathbb C^{N_T\times12}$ denote the noisy first-RB observation. Its thin SVD is $\mathbf Y_0=\mathbf U\mathbf\Sigma\mathbf V^H$. With $q=12$, we set
\begin{align}
\mathbf U_T^{(0)}[:,1:q]&=\mathbf U,\\
\mathbf U_F^{(0)}[\mathcal I_0,1:q]&=\mathbf V^*,\\
\mathbf U_t^{(0)}[0,1:q]&=\operatorname{diag}(\mathbf\Sigma).
\end{align}
The first $q$ frequency columns are zero outside $\mathcal I_0$. The remaining $R-q$ spatial and frequency columns are independent complex Gaussian directions normalized to unit norm. Their temporal coefficients have independent random phases and magnitude $0.01\sqrt{q^{-1}\sum_{r=1}^{q}\Sigma_{rr}^2}$. Thus every component has nonzero spatial and frequency directions, even though individual unobserved entries of the SVD columns remain zero. The ordinary online update starts at the first frame. At 10/20 dB this initializer uses the raw target first RB; at 0/5 dB it uses the spatially denoised observation specified in Section~\ref{sec:low_snr}.

\subsection{Statistical Initialization with Additional Dispersed Pilots}
The second initializer retains all 12 first-RB tones and acquires $m$ additional real pilot tones outside that RB. For $m\geq2$, their zero-based indices are obtained by rounding $m$ equally spaced points between 12 and 575. The main comparison uses $m=6$, giving 18 first-frame pilots. Every subsequent frame uses the same 12-tone periodic RB scan as the direct initializer. Over 1000 frames, this changes the measured pilot-tone count from 12,000 to 12,006, an increase of $0.05\%$; the first-frame count and energy at fixed per-pilot SNR increase by $50\%$.

Training trajectories provide a 64-dimensional spatial subspace $\mathbf Q$ and a delay dictionary $\mathbf D\in\mathbb C^{576\times256}$. We form these statistics from seven other users, using 16 snapshots per user at frame indices $0,4,\ldots,60$. Each snapshot is normalized by its element RMS. The leading eigenvectors of the mean spatial covariance form $\mathbf Q$. An oversampled 2304-point inverse Fourier transform provides an average delay-power profile; its 256 strongest bins define $\mathbf D$ and normalized initial powers $\boldsymbol\gamma_0$. These statistics are prepared independently at each speed.

For startup indices $\Omega_0$, the spatially projected observation is $\widetilde{\mathbf Y}_0=\mathbf Q\mathbf Q^H\mathbf Y_0$. After RMS normalization, transpose it to $\mathbf Z\in\mathbb C^{|\Omega_0|\times N_T}$ and write $\mathbf D_\Omega=\mathbf D[\Omega_0,:]$. The common-delay support is modeled through shared row variances, following the multiple-measurement-vector sparse Bayesian principle \cite{zhang2011sbl}. Our implementation uses conditionally independent measurement vectors and a training-power anchor. With $\boldsymbol\Gamma=\operatorname{diag}(\boldsymbol\gamma)$, its updates are
\begin{align}
\mathbf C_y&=\mathbf D_\Omega\boldsymbol\Gamma\mathbf D_\Omega^H+\nu\mathbf I,\\
\mathbf M&=\boldsymbol\Gamma\mathbf D_\Omega^H\mathbf C_y^{-1}\mathbf Z,\\
v_\ell&=\gamma_\ell-\gamma_\ell^2
 [\mathbf D_\Omega^H\mathbf C_y^{-1}\mathbf D_\Omega]_{\ell\ell},\\
\gamma_\ell^+&=(1-a)\left(N_T^{-1}\|\mathbf M_{\ell,:}\|_2^2+v_\ell\right)
 +a\gamma_{0,\ell}.
\end{align}
Twenty-four iterations yield $\widehat{\mathbf H}_0=(\mathbf D\mathbf M)^T$, with the observation scale restored. The anchor $a\in\{0.05,0.25\}$ and normalized noise parameter $\nu\in\{0.003,0.01,0.03,0.1\}$ are selected using only validation first-frame mean NMSE in dB, separately for each speed and SNR. 

We factor $\widehat{\mathbf H}_0$ by SVD and retain singular components above $10^{-6}$ times its leading singular value, up to width 40. The remaining spatial and frequency directions are obtained by projecting the training subspace and delay dictionary away from the retained directions and taking orthonormal bases. Each complementary spatial and frequency column is scaled by $0.1$, and its temporal coefficient is $0.01\sigma_1$. Consequently, each added rank-one term has amplitude $10^{-4}\sigma_1$. This small completion preserves the initial estimate while keeping every component learnable. The first output is this completed estimate; the usual online updates begin at frame 2. Actual startup samples enter the spatial and temporal histories. The complete first RB also enters its frequency-block history.

\section{Simulation Setup and Results Analysis}

We evaluate the complete CP-PBCD tracker on time-varying channels generated with a 3GPP-compliant model. The main comparison covers three low-mobility speeds under periodically swept RB sampling.

\subsection{Simulation Environment and Parameter Settings}
\label{sec:simulation_settings}

The simulation data are generated using the QuaDRiGa platform in accordance with the 3GPP TR 38.901 standard \cite{jaeckel2014quadriga,ademaj20163gpp}. We consider an urban macro non-line-of-sight (UMa-NLOS) scenario, where the user equipment (UE) moves along a straight trajectory with different constant speeds to emulate different levels of Doppler dynamics.

The receiver observes one RB every 1 ms and periodically sweeps the band. Table~\ref{tab:sim_params} lists the simulation parameters.

\begin{table}[t]
\centering
\caption{Simulation parameter settings}
\label{tab:sim_params}
\renewcommand{\arraystretch}{1.08}
\setlength{\tabcolsep}{4pt}
\footnotesize
\begin{tabularx}{\linewidth}{>{\raggedright\arraybackslash}p{0.43\linewidth} >{\raggedright\arraybackslash}X}
\toprule
\textbf{Category} & \textbf{Value} \\
\midrule

\multicolumn{2}{l}{\textit{System parameters}} \\
Center frequency & 3.5 GHz \\
System bandwidth & 20 MHz \\
Subcarrier spacing (SCS) & 30 kHz \\
Active subcarriers $N_F$ & 576 (48 RBs) \\
Base station array & 64 ports (32 dual-polarized positions) \\
User array & 4-element patch model \\
Stacked spatial dimension $N_T$ & $64\times4=256$ \\
Total frames $N_t$ & 1000 \\

\addlinespace[2pt]
\multicolumn{2}{l}{\textit{Channel dynamics}} \\
Primary low-speed case & 3.6 km/h \\
Additional low-speed cases & 5.4 and 7.2 km/h \\

\addlinespace[2pt]
\multicolumn{2}{l}{\textit{Sampling scheme}} \\
Gap = 1 ms & 1 RB per 1 ms ($s=12$) \\

\addlinespace[2pt]
\multicolumn{2}{l}{\textit{Hyperparameters}} \\
Initial and solver CP width & 40 \\
Allowed active-component range & 12--40 \\
Spatial history & 48 frames \\
Frequency history & 3 visits per RB \\
Temporal history & 24 frames \\
Base $\tau_F,\tau_t$ & $1,10^{5}$ \\
Base $\gamma_T,\gamma_F,\gamma_t$ & $1,2,500$ \\
Temporal solver & GPU, two PCG iterations \\
Online update budget & One sweep per frame \\

\bottomrule
\end{tabularx}
\end{table}

The main evaluation uses UE11--UE20 at 3.6, 5.4, and 7.2 km/h. UE01--UE07 provide statistical priors and UE08--UE10 select startup and physical-model settings. The four cache-based baseline settings remain frozen from their UE07--UE08 validation. Each speed uses distinct propagation realizations. All methods receive identical noisy observations for a given user and SNR.

All CP solvers allocate $R=40$ columns and perform one update sweep per frame. First-RB completion and the online CNN use two seeds, averaged within user; statistical startup is deterministic. CP first-RB acquires 12 pilots at every frame. CP +6 acquires 18 at the first frame and the same 12 as all other methods thereafter, adding six measurements over the 1000-frame trajectory.

\subsection{Low-SNR Spatial Denoising}
\label{sec:low_snr}
At 0/5 dB, both CP versions project each arriving observation onto the 64-dimensional training-user subspace, $\widetilde{\mathbf S}_t=\mathbf Q\mathbf Q^H\mathbf S_t$. The speed-specific $\mathbf Q$ is constructed as in Section~\ref{sec:initialization}. First-RB SVD uses this projected block; CP +6 retains its statistical startup estimate. Subsequent updates use the same projected observations and two-PCG-iteration budget.

Validation-selected multipliers for $(\gamma_T,\gamma_F,\gamma_t,\tau_F,\tau_t)$ are $(256,1,256,256,256)$ at 0 dB and $(4,1,1,16,1)$ at 5 dB. They are fixed across the three speeds. At 10/20 dB the input is unprojected and all multipliers equal one. Spatial projection and stronger regularization suppress noise without adding solver iterations.

\subsection{Channel Degradation Models and Evaluation Metrics}

\subsubsection{Noise Modeling}

The observation model includes additive white Gaussian noise (AWGN),
\[
\mathbf{E}_t \sim \mathcal{CN}(\mathbf{0}, \sigma^2 \mathbf{I}),
\]
where the noise variance $\sigma^2$ is determined by the target signal-to-noise ratio (SNR).

\subsubsection{Evaluation Metrics}

We measure reconstruction error with NMSE and directional agreement with correlation.

The normalized mean squared error (NMSE) is
\[
\mathrm{NMSE}(t)
=
\frac{\|\mathbf{h}_t-\hat{\mathbf{h}}_t\|_F^2}{\|\mathbf{h}_t\|_F^2},
\]
and is reported as $10\log_{10}\mathrm{NMSE}(t)$. At each frame, direct first-RB CP-PBCD and CNN metrics are averaged over two seeds within each UE, then over ten users. Statistical initialization is deterministic and is averaged over those same ten users. Full-trajectory scores average these dB values over 1000 frames. Between-user population standard deviations use the UE as the statistical unit and describe variability in the reported mean comparisons.

The normalized correlation between the estimated and true channels is
\[
\rho(t)
=
\frac{|\langle \mathbf{h}_t,\hat{\mathbf{h}}_t\rangle|}{\|\mathbf{h}_t\|\,\|\hat{\mathbf{h}}_t\|}.
\]
Correlation captures agreement in beam and phase structure and is relevant to beamforming quality.

\subsection{Baseline Implementations}
\label{sec:baseline_implementation}
We implement four causal cache-based baselines, including Fourier-slice nuclear shrinkage inspired by~\cite{lu2016trpca} and neural frequency reconstruction inspired by~\cite{soltani2019channel}. Every method starts without target history, uses the same noisy RB observations, and produces all 576 subcarriers after each observation. Unseen cache entries are zero.

\emph{CZOH:} Each arriving 12-tone block overwrites its cached values. All other frequencies retain their most recent observations; there is no interpolation or denoising.

\emph{RB-AR:} For each RB, store its latest three visits, separated by 48 frames. With their vectorized blocks $\boldsymbol y_1,\boldsymbol y_2,\boldsymbol y_3$, fit a shared two-tap coefficient vector
\[
\boldsymbol p=(\mathbf X^H\mathbf X+10^{-2}\mathbf I_2)^{-1}\mathbf X^H\boldsymbol y_3,
\quad\mathbf X=[\boldsymbol y_1,\boldsymbol y_2].
\]
The next-visit forecast is $[\boldsymbol y_2,\boldsymbol y_3]\boldsymbol p$, with each complex entry magnitude clipped to 20 in the normalized channel units. Before three visits, the forecast equals the latest block. Between visits, linearly interpolate from that block to its forecast using the age fraction $\min\{\mathrm{age}/48,1\}$. The current block is exactly its new observation.

\emph{TNN-cache:} Divide the cached matrix by its RMS over observed entries. Reshape it as a $256\times12\times48$ tensor, with the last axis indexing RBs. Set $w_f=2^{-\mathrm{age}_f/24}$ for seen frequencies and zero otherwise. Starting from the normalized cache $Y$, take eight proximal-gradient steps
\[
X^{j+1}=\mathcal S_{30}\!\left(X^j-W\odot(X^j-Y)\right),\quad X^0=Y.
\]
Here $\mathcal S_{30}$ takes a length-48 FFT on the RB axis, applies singular-value soft thresholding at 30 to each $256\times12$ Fourier slice, and applies the inverse FFT. This gives a unit-step weighted cache fit with a Fourier-slice nuclear penalty. Restore the cache RMS after eight steps; there is no convergence-based stopping rule.

\emph{Online CNN:} Inspired by neural channel reconstruction~\cite{soltani2019channel}, we implement a shared per-spatial-channel 1D frequency network. It receives four channels: normalized cache real and imaginary parts, age clipped to $[0,1]$ after division by 48, and a seen-frequency mask. Four convolution/GELU blocks have width 16, kernel size 5, and dilations $1,4,16,64$, followed by a kernel-1 convolution with two output channels. Zero initialization of the output layer makes the initial correction zero. Its prediction is the complex cache plus the age-gated network residual. Before ingesting the current block, perform one Adam step against that block's noisy real/imaginary observations, using the mean squared error over its $256\times2\times12$ entries. The training input contains only previous observations. Then ingest the new block and evaluate the full band; the age-zero block is set exactly to its observation. Adam uses $(\beta_1,\beta_2)=(0.9,0.999)$, $\epsilon=10^{-8}$, no weight decay, and learning rates $0.003,0.003,0.001$ at the three increasing speeds. There is no offline checkpoint. Results are averaged over two random initializations.

These configurations were frozen on the earlier UE07--UE08 validation cohort; statistical startup selection uses UE08--UE10. Direct CP completion is evaluated over two random initializations. Noise is shared across methods, with seed $40000+100\,\mathrm{SNR}_{\mathrm{dB}}+u$ for user index $u$. For each frame, its simulated noise variance is the full-band mean channel energy over nonzero entries divided by the linear SNR. True channels determine simulated noise and evaluation metrics, not estimator inputs.

\subsubsection{Physical Angle--Delay Adaptations}
\label{sec:physical_baselines}
\emph{Phys-Fixed} and \emph{Phys-Refit} adapt the acquisition-and-tracking architecture of Wan and Liu~\cite{wan2024twostage} to our 12-tone RB stream. Both estimate an initial model from the first RB and rebuild it at frame 48, when a full frequency sweep becomes available. Phys-Fixed then freezes its geometry. Phys-Refit re-estimates geometry at frames $96,144,\ldots$, using the latest 96 arrived observations. Their coefficient tracker is the Gaussian update specified below, replacing the original dynamic turbo/EM procedure.

The array has eight polarization/receive-port groups with a $4\times8$ aperture per group. Shared delays, directions, and Doppler frequencies define path atoms
\[
\boldsymbol b_k(n)=\boldsymbol a(u_k,v_k)e^{-\mathrm j2\pi f_n\tau_k}
e^{\mathrm j2\pi\nu_k(t_n-t_{\rm end})},
\]
with group-specific complex gains. We fit 128 paths, with delays in $[0,6]~\mu$s, $u_k^2+v_k^2<1$, and $|\nu_k|\leq11.675(v/3.6)$ Hz at speed $v$ in km/h. Residual correlation over a 1537-point delay grid on $[-0.1,6]~\mu$s and a $48\times48$ direction grid initializes bounded continuous fitting. Only measured tones enter the loss. L-BFGS uses up to 20 iterations per added path and 100 after every eighth addition. At frame 48, a first-sweep seed is refined jointly using the 48 actual observation times and a 300-iteration budget. Later periodic refits use 80 iterations from the preceding model.

Complex gains are eliminated by ridge least squares with coefficients $10^{-7}N$ for seeding and $10^{-6}N$ for joint fitting, where $N$ counts observed tones. The loss is normalized by observation energy. Strong-Wolfe line search uses initial steps 0.7 and 0.5 for seeding and joint fitting, respectively, with objective-change tolerance $10^{-9}$.

An SVD orthonormalizes the full-band path dictionary, retaining directions above $10^{-3}$ of its largest singular value. Gaussian random-walk tracking updates the eight groups' coefficients from each current RB. Initial and process covariances are $0.01s_z\mathbf I$ and $10^{-4}s_z\mathbf I$, where $s_z$ is initial coefficient energy. Noise variance is observation energy divided by $1+10^{\mathrm{SNR}/10}$, using frame 1 initially and the first 48 frames after rebuilding. Refits rebuild both basis and covariance. Parameters are frozen from UE08--UE10.

Ordinary coefficient updates use CUDA Graph replay. Joint fitting and tracking use complex double precision, with single-precision seeding. A separable spatial/frequency--Doppler Gram computation reduces joint-solver cost. This implementation compares continuous physical-model adaptation with incremental CP factor updates under the same causal observations.

\subsection{Current Full-Band Reconstruction}
\begin{figure*}[!t]
\centering
\includegraphics[width=\linewidth]{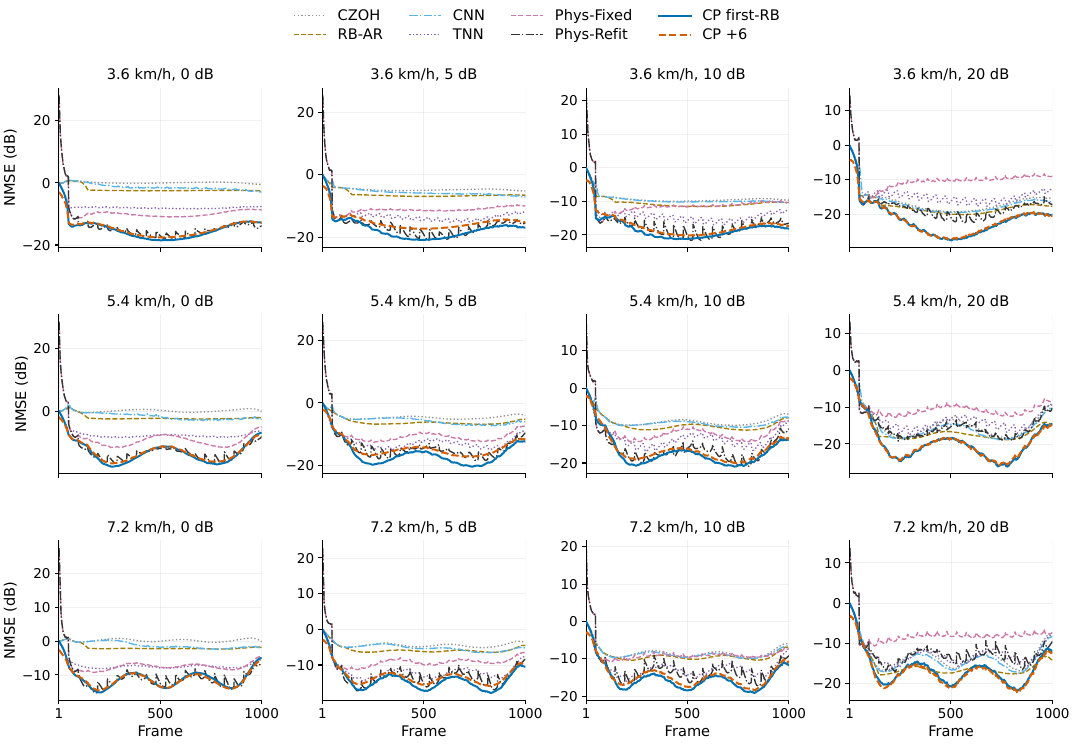}
\caption{Current-channel NMSE over complete 1000-frame trajectories: three speeds, four SNRs, ten test users, and eight methods. Both physical methods start from the first RB and rebuild at frame 48; Phys-Refit subsequently refits every 48 frames. CP +6 adds six real startup tones. Both CP versions use spatial denoising at 0/5 dB. Curves average per-user dB values without temporal smoothing.}
\label{fig:main_all}
\end{figure*}

Figure~\ref{fig:main_all} compares all eight methods on shared axes. Table~\ref{tab:main_all} averages per-user dB values over the complete frames 1--1000, including startup. Both physical adaptations output from frame 1 and rebuild at frame 48; Phys-Fixed then freezes geometry, whereas Phys-Refit continues to rebuild every 48 frames. The two CP versions differ in their first-frame acquisition and initializer.

\begin{table*}[!t]
\centering\footnotesize
\caption{Mean NMSE (dB) over complete frames 1--1000 and reference online computation}
\label{tab:main_all}
\setlength{\tabcolsep}{4.5pt}
\begin{tabular}{rrrrrrrrrr}
\toprule
Speed & SNR & CZOH & RB-AR & CNN & TNN & Phys-Fixed & Phys-Refit & CP first-RB & CP +6 \\
\midrule
3.6 & 0 & 0.08 & -2.10 & -1.32 & -7.80 & -9.19 & -13.95 & \textbf{-14.93} & -14.59 \\
3.6 & 5 & -4.68 & -6.30 & -5.57 & -13.21 & -10.48 & -15.92 & \textbf{-17.35} & -15.16 \\
3.6 & 10 & -9.21 & -10.39 & -9.48 & -14.25 & -10.87 & -16.77 & \textbf{-18.03} & -17.26 \\
3.6 & 20 & -16.38 & -17.52 & -16.70 & -14.56 & -9.97 & -17.58 & \textbf{-21.49} & \textbf{-21.49} \\
\midrule
5.4 & 0 & 0.14 & -2.03 & -1.57 & -7.48 & -8.44 & -12.11 & \textbf{-13.00} & -12.71 \\
5.4 & 5 & -4.48 & -6.02 & -5.57 & -12.36 & -9.99 & -13.71 & \textbf{-16.13} & -14.30 \\
5.4 & 10 & -8.68 & -9.74 & -9.04 & -13.25 & -11.59 & -15.76 & \textbf{-16.94} & -16.23 \\
5.4 & 20 & -14.63 & -16.27 & -15.34 & -13.52 & -10.28 & -14.90 & -19.53 & \textbf{-19.55} \\
\midrule
7.2 & 0 & 0.20 & -1.92 & -1.17 & -7.26 & -6.94 & -10.47 & -10.97 & \textbf{-11.14} \\
7.2 & 5 & -4.33 & -5.70 & -5.07 & -11.69 & -8.65 & -12.61 & \textbf{-14.03} & -13.05 \\
7.2 & 10 & -8.30 & -9.10 & -8.56 & -12.44 & -8.54 & -12.70 & \textbf{-15.10} & -14.47 \\
7.2 & 20 & -13.19 & -15.50 & -13.81 & -12.66 & -7.82 & -12.28 & -16.44 & \textbf{-16.96} \\
\midrule
\multicolumn{2}{l}{Mean time (ms/frame)} & 0.122 & 0.198 & 2.496 & 3.694 & 2.558 & 25.167 & 0.717 & 0.674 \\
\bottomrule
\end{tabular}
\par\vspace{3pt}\parbox{0.98\linewidth}{\scriptsize Speed is in km/h and pilot SNR in dB. NMSE averages the ten test users after within-user seed averaging, including all startup frames. Runtime uses UE11--UE12 at 3.6 km/h and 20 dB, frames 97--1000; periodic refits are included and one-time setup is separate. CP +6 acquires 18 first-frame tones; all other frames and methods use 12.}
\end{table*}
First-RB CP-PBCD achieves lower complete-trajectory mean NMSE than all six baselines in every speed/SNR condition. At 3.6 km/h and 20 dB, both CP versions reach approximately $-21.49$ dB, compared with $-9.97$ dB for Phys-Fixed and $-17.58$ dB for Phys-Refit. First-RB CP-PBCD improves on periodic refitting by 3.91 dB. At 5.4 and 7.2 km/h and 20 dB, its gains over refitting are 4.63 and 4.16 dB, respectively.

The physical adaptations expose the cost of maintaining a path model under narrow-RB acquisition. Rebuilding once after a complete scan improves startup geometry, but subsequent fixed-geometry tracking accumulates mismatch. Periodic refitting improves its accuracy at all twelve conditions and creates visible reset transients. CP-PBCD continuously adapts its factors without recurring joint path estimation, obtaining lower full-trajectory error with a smaller update budget.

Spatial denoising strengthens the low-SNR results. At 0 and 5 dB, first-RB CP-PBCD outperforms all six baselines at every speed. Additional startup tones improve early frequency coverage, while long-run accuracy is governed by the subsequent denoising and factor updates. The controlled study below isolates startup-budget effects with a common statistical initializer.

\subsection{Online Computation and Startup Costs}
\label{sec:runtime}
All methods run on an NVIDIA RTX 4080 SUPER with an Intel Core i7-14700KF host. The reference online workload uses UE11--UE12 at 3.6 km/h and 20 dB, frames 97--1000. The six existing implementations use their three serial timing repetitions; the two physical adaptations use the completed serial runs under the updated startup protocol. Timing includes pilot transfer, updates, full-band synthesis, output transfer to pinned host memory, and GPU synchronization. Phys-Refit timings include nonlinear refitting, basis reconstruction, and graph rebuilding. Data loading, scoring, and file output are excluded.

CP first-RB and CP +6 average $0.717$ and $0.674$ ms per online update, below the 1-ms observation interval. Phys-Fixed and Phys-Refit require $2.558$ and $25.167$ ms, respectively. Incremental CP updates provide a $35.1\times$ speed advantage over periodic refitting while attaining lower reconstruction error. CUDA Graph replay, two temporal PCG iterations, and parallel frequency solves keep the online work bounded; component checks remain included.

\begin{table}[!t]
\centering\footnotesize
\caption{One-time computation in seconds, averaged over twelve conditions and ten users}
\label{tab:setup}
\begin{tabular}{lrr}
\toprule
Method & First output & Frame-48 rebuild \\
\midrule
CZOH & 0.063 & --- \\
RB-AR & 0.068 & --- \\
CNN & 0.016 & --- \\
TNN & 0.036 & --- \\
Phys-Fixed & 24.868 & 20.194 \\
Phys-Refit & 24.863 & 20.164 \\
CP first-RB & 0.091 & --- \\
CP +6 & 2.061 & --- \\
\bottomrule
\end{tabular}
\end{table}
Table~\ref{tab:setup} separates first-output computation from the frame-48 rebuild. All methods use the first arriving RB without additional observation waiting. First-output time includes model construction, initialization, first-shape graph preparation, and output synthesis. Training-prior construction and CUDA context creation occur beforehand. These are one-time setup costs, separate from the online update times in Table~\ref{tab:main_all}.

\subsection{Avoiding an All-Zero Component State}
In a CP model, the contribution of component $r$ is $\mathbf a_r\circ\mathbf b_r\circ\mathbf c_r$. If all three factors of this component start at zero, the alternating updates have no nonzero counterpart with which to estimate a new direction. A restoration mechanism that saves an already-zero direction cannot correct this state. This differs from an individual zero entry in an otherwise nonzero component, which can be updated when observations provide information about that entry.

\begin{figure}[H]
\centering
\includegraphics[width=\linewidth]{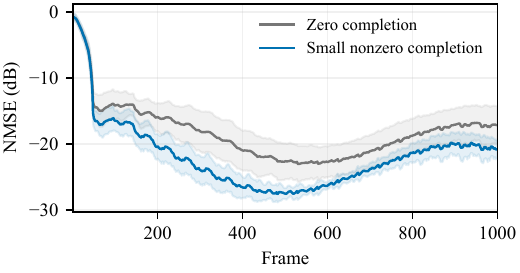}
\caption{Paired first-RB initialization control at 3.6 km/h and 20 dB with identical observations and statistical estimates. Only weak-component completion changes. Lines and shading show ten-user mean NMSE and one standard deviation.}
\label{fig:startup_nonzero}
\end{figure}

We isolate this issue using identical first-RB observations and the same statistical estimate, changing only the completion of weak components. Figure~\ref{fig:startup_nonzero} compares the resulting trajectories.

 Over ten test users at 3.6 km/h and 20 dB, small nonzero completion improves full-trajectory mean NMSE from $-18.179$ to $-21.839$ dB; the last-100-frame mean improves from $-17.082$ to $-20.364$ dB. All ten users improve in full-trajectory mean NMSE. The initialization perturbation is at most $0.0562\%$ in relative Frobenius norm. All 40 columns start learnable, and 32--37 remain active at the end in this first-RB statistical configuration. Preserving weak directions allows later measurements to activate them before residual-based component selection.

\subsection{Effect of the Number of Additional Startup Pilots}
\label{sec:pilot_budget}
\begin{figure}[H]
\centering
\includegraphics[width=\linewidth]{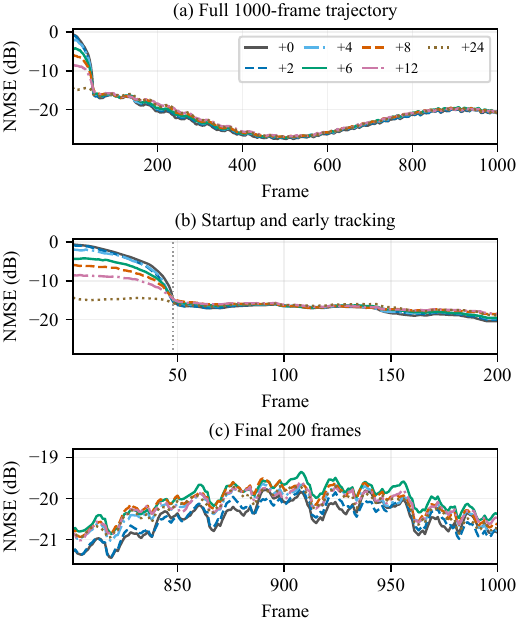}
\caption{First-RB 12 pilots plus $m$ extra dispersed startup pilots, with $m=0,2,4,6,8,12,24$. Only the first frame differs. Subsequent frames use identical 12-pilot RB scans and the same CP-PBCD updates. Curves average per-user NMSE in dB over UE11--UE20 at 3.6 km/h and 20 dB, without temporal smoothing. All configurations use small nonzero completion. Panels (a)--(c) show the full trajectory, early acquisition, and late tracking; legend entries denote the number of extra tones.}
\label{fig:startup_budget}
\end{figure}

Using the statistical initializer and nonzero completion, we test $m\in\{0,2,4,6,8,12,24\}$. Every configuration retains the first RB, so the total startup budget is $12+m$. The same eight-candidate first-frame validation procedure selects the Bayesian parameters separately for each budget. All later noisy observations and update rules are identical. Figure~\ref{fig:startup_budget} and Table~\ref{tab:startup_budget} compare startup coverage within this common initializer family and selection protocol.

\begin{table}[!htb]
\centering\footnotesize
\caption{Startup-budget study: mean NMSE in dB over ten users}
\label{tab:startup_budget}
\setlength{\tabcolsep}{4pt}
\begin{tabular}{rrrrr}
\toprule
Extra $m$ & Total & Frame 1 & Frame 48 & Last 100 \\
\midrule
0 & 12 & -0.646 & -15.203 & -20.364 \\
2 & 14 & -0.878 & -15.463 & -20.352 \\
4 & 16 & -1.939 & -14.802 & -20.157 \\
6 & 18 & -4.347 & -14.793 & -19.937 \\
8 & 20 & -5.948 & -14.433 & -20.092 \\
12 & 24 & -8.567 & -14.668 & -20.114 \\
24 & 36 & -14.237 & -15.670 & -20.202 \\
\bottomrule
\end{tabular}
\end{table}

\begin{figure*}[t]
\centering
\includegraphics[width=0.99\linewidth]{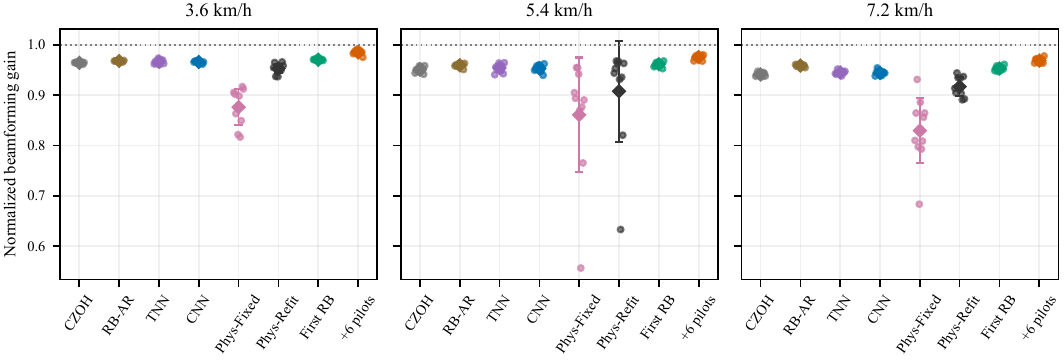}
\caption{Normalized single-stream beamforming gain for eight methods at pilot SNR 20 dB. Points show user means over all 1000 frames and 576 subcarriers; diamonds and error bars show the ten-user mean and population standard deviation. Physical methods use the main-experiment startup and refitting protocols; CP versions retain their stated startup budgets.}
\label{fig:beamforming_gain}
\end{figure*}
Additional dispersed pilots improve first-frame NMSE from $-0.646$ to $-14.237$ dB as $m$ increases from zero to 24. The last-100-frame means occupy a much smaller $0.427$-dB range. Better startup coverage therefore primarily shortens early acquisition, particularly before the first 48-RB sweep is complete. With nonzero component directions available, incremental updates attain similar late tracking accuracy across these initial estimates. This behavior agrees with the decaying initial-error term in the local tracking bound.

\subsection{Beamforming Performance}
\label{sec:beamforming}
Figure~\ref{fig:beamforming_gain} compares the beamforming gains. At pilot SNR 20 dB, we evaluate single-user, single-stream fully digital beamforming over all 1000 frames and 576 subcarriers of each test user. Each stacked estimate is restored to a $4\times64$ channel matrix. Its dominant left and right singular vectors, $\mathbf w$ and $\mathbf v$, provide the combiner and precoder. They depend only on the estimate. The normalized gain is
\begin{equation}
\eta^{\mathrm{BF}}_{u,f,t}=\frac{|\mathbf w_{u,f,t}^{H}\mathbf H_{u,f,t}\mathbf v_{u,f,t}|^2}{\sigma_1^2(\mathbf H_{u,f,t})}.
\end{equation}
For a zero estimate we use fixed first-coordinate beams. We average gains over frames and subcarriers, then over seeds where applicable, then over users. The true channel is used only to evaluate the gain and its ideal normalization.

The direct first-RB version obtains mean normalized gains of 97.04\%, 96.16\%, 95.30\% at increasing speeds. The six-extra-pilot version obtains 98.50\%, 97.58\%, 96.88\%. Phys-Fixed obtains 87.63\%, 86.09\%, and 82.96\%; Phys-Refit obtains 95.27\%, 90.79\%, and 91.69\%. Both CP versions retain higher mean beamforming gain than the physical adaptations at all three speeds.

Single-stream spectral efficiency is $\log_2(1+\rho_{\mathrm d}g_{u,f,t}/q_u)$ under white Gaussian noise, Gaussian signaling, and receiver knowledge of the effective scalar channel. Here $g_{u,f,t}=|\mathbf w^H\mathbf H\mathbf v|^2$, and $q_u$ is the user's mean true-channel element energy, used only for evaluation. At data SNR $\rho_{\mathrm d}=10$ dB, first-RB and CP +6 attain respectively $(10.9116,10.8062,10.8621)$ and $(11.0398,10.9535,11.0036)$ bit/s/Hz across the three speeds, before pilot-overhead deduction.

\subsection{Two-by-Two Smoothness Ablation}
\begin{figure}[H]
\centering
\includegraphics[width=\linewidth]{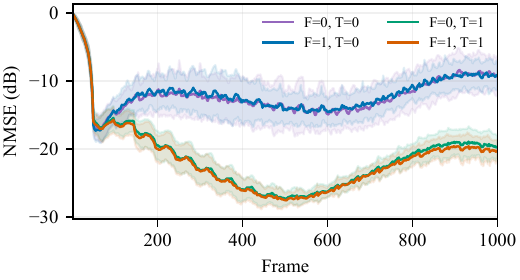}
\caption{Two-by-two F/T ablation with direct first-RB initialization. Lines show ten-user mean NMSE in dB; shading shows one between-user population standard deviation after within-user averaging over two completion seeds.}
\label{fig:component_ablation}
\end{figure}

Table~\ref{tab:component_ablation} and Fig.~\ref{fig:component_ablation} toggle frequency (F) and explicit temporal (T) smoothness at 3.6 km/h and 20 dB for UE11--UE20. All four configurations share direct first-RB initialization, noisy pilots, age-weighted histories, component management, width 40, and two temporal PCG iterations. Previous-row initialization and the temporal proximal term remain active when T is disabled. Results average two completion seeds within each user.

\begin{table}[H]
\centering\footnotesize
\caption{F/T ablation: ten-user mean $\pm$ standard deviation}
\label{tab:component_ablation}
\setlength{\tabcolsep}{4pt}
\begin{tabular}{ccrr}
\toprule
F & T & NMSE (dB) & Correlation \\
\midrule
0 & 0 & $-12.08\pm1.88$ & $0.9507\pm0.0170$ \\
1 & 0 & $-11.91\pm1.05$ & $0.9522\pm0.0091$ \\
0 & 1 & $-21.16\pm0.92$ & $0.9804\pm0.0029$ \\
1 & 1 & $-21.49\pm0.70$ & $0.9809\pm0.0028$ \\
\bottomrule
\end{tabular}
\end{table}

Adding temporal regularization with F disabled improves mean NMSE by 9.07 dB, from $-12.08$ to $-21.16$ dB. Adding F with T enabled provides a further 0.33 dB. The complete F+T configuration reaches $-21.49\pm0.70$ dB, identifying temporal coupling as the dominant smoothness contribution.

\section{Conclusion}
CP-PBCD reconstructs current full-band MIMO channels from 1-ms periodically swept RB observations through age-weighted histories, local regularization, and adaptive component management. Its fixed-budget GPU updates provide submillisecond average online computation on the reference workload. First-RB CP-PBCD achieves lower complete-trajectory mean NMSE than six baselines across all twelve speed/SNR conditions. At 3.6 km/h and 20 dB, it improves on periodic physical refitting by 3.91 dB with a $35.1\times$ speed advantage. Nonzero completion supports continued factor learning, while dispersed startup pilots accelerate acquisition with little change in the controlled study's late tracking accuracy.

\bibliographystyle{IEEEtran}
\bibliography{references}

\end{document}